\documentclass[pre,twocolumn]{revtex4}
\usepackage[dvips]{graphicx}
\usepackage{amsmath,axodraw}

\newcommand{\avg}[1]{\ensuremath{\langle #1 \rangle}}

\begin{document}

\title{Drowsy Cheetah Hunting Antelopes: A Diffusing Predator Seeking
Fleeing Prey}
\date{\today}

\author{Karen Winkler}
\author{Alan J.\ Bray}
\affiliation{Department of Physics and Astronomy, University of 
Manchester, Manchester M13 9PL, U.K.}
 
\begin{abstract}
We consider a  system of three random walkers  (a `cheetah' surrounded
by two  `antelopes') diffusing in  one dimension. The cheetah  and the
antelopes  diffuse,  but  the   antelopes  experience  in  addition  a
deterministic  relative   drift  velocity,  away   from  the  cheetah,
proportional to their  distance from the cheetah, such  that they tend
to  move  away from  the  cheetah  with  increasing time.   Using  the
backward Fokker-Planck  equation we calculate, as a  function of their
initial  separations,  the probability  that  the  cheetah has  caught
neither antelope after infinite time.
\end{abstract}

\maketitle


\section{Introduction}
Diffusion controlled reactions of three particles in one dimension can
be completely understood by mapping  the process to a single diffusing
particle in a  two-dimensional wedge \cite{fishergelfand, benavraham},
where the lines of reaction  -- the positions where two particles meet
-- correspond to the boundaries of  the wedge.  By this elegant method
Fisher   and  Gelfand   \cite{fishergelfand}   investigated  diffusing
particles termed  vicious walkers  which annihilate on  meeting, while
Redner   and  Krapivsky   \cite{krapivskyredner96,  rednerkrapivsky99}
studied the equivalent capture reaction, where a single diffusing prey
(`lamb')  is eliminated  on  meeting one  of  two diffusing  predators
(`lions').  One of  the main properties of interest  in these problems
is  the  survival  probability   of  all  three  vicious  walkers  or,
equivalently, the single prey.

In this  paper we introduce a  three-particle system  in one dimension
consisting  of two prey (`antelopes'),  surrounding a  single predator 
(`cheetah'). So far this is just another statement of the vicious walker 
problem with three walkers. Our model differs from the standard model, 
however,  as follows. Besides  performing  a  diffusive motion  
all particles are subjected to a drift which increases linearly with 
their position coordinate. Considering the  case where both species have 
the same  diffusion constant, the equation of motion for  the antelopes
($A_{1}$,  $A_{2}$)  and  the  cheetah ($C$)  with  initial  positions
$x_{A_{1}}<x_{C}<x_{A_{2}}$ is taken to be:
\begin{equation}
\dot{x}_i = a x_i + \eta_i(t),\quad i=A_{1},A_{2},C
\label{langevin}
\end{equation}
where $a$ is the strength of the drift. The  Langevin noise $\eta_i(t)$ 
is a Gaussian white  noise with mean zero and correlator
\begin{equation}
\avg{\eta_i(t) \eta_j(t^\prime)} = 2 D \delta_{ij} \delta(t-t^\prime).
\label{noise}
\end{equation}
Equation  (\ref{langevin})  models  the  overdamped  motion  of  three
particles moving independently in an inverted parabolic potential. The
calculation  of  the  time-dependent  survival probability  for  three
vicious walkers in a conventional parabolic potential (i.e.\ with $a<0$
in Eq.\ (\ref{langevin})) has been presented elsewhere \cite{BW}.

Studying   the   problem    in   the   {\em   relative}   coordinates,
$y_{1}=x_{C}-x_{A_{1}}$ and  $y_{2}=x_{A_{2}}-x_{C}$, the equations of
motion have  terms linearly  depending on these  relative coordinates.
Therefore  the antelopes are  always drifting  away from  the cheetah,
with a drift rate proportional  to the distance from the predator.  As
a result,  there is a  nonzero probability that both  antelopes wander
off  to infinity  without meeting  the cheetah  if they  are initially
separated from  the cheetah.   Defining the process  to be  `alive' if
neither  of the  antelopes  has met  the  cheetah, we  find a  nonzero
survival probability $Q(y_{1},y_{2})$ for $y_{1},y_{2}>0$.  The aim of
this    paper   is   to    calculate   this    survival   probability,
$Q(y_{1},y_{2})$,  in  the limit  of  infinite  time,  given that  the
antelopes started initially at  relative distances $y_{1}$ and $y_{2}$
from the cheetah.

To provide  context for our result  we consider first a  cheetah and a
single antelope.  In  section III the case of  a cheetah surrounded by
two antelopes is investigated by mapping the process to a single diffusing
particle in a two-dimensional wedge. Section IV is a short conclusion.

\section{A cheetah and a single antelope}
\label{simpleexample}
The  dynamics of  a  cheetah  ($C$) and  an  antelope ($A_{1}=A$)  is
described  by  the   Langevin  equation  (\ref{langevin})  with  noise
correlator (\ref{noise}). The process  terminates when the cheetah and
the  antelope  meet, i.e.\  when  $x_{A}=x_{C}$.  Setting the  initial
positions  as  $x_{A} <  x_{C}$  we  introduce  a relative  coordinate
$y_{1}=y=x_{C}-x_{A}$ which obeys the Langevin equation:
\begin{equation}
\dot{y} = a y + \xi(t),
\end{equation}
where $\xi(t)=\eta_{C}-\eta_{A}$ is a Gaussian white noise with mean zero 
and correlator:
\begin{equation}
\avg{\xi(t) \xi(t^\prime)} = 4 D \delta(t-t^\prime).
\end{equation}
The probability $Q(y)$ that the  antelope has survived in the limit of
infinite time, given  that antelope and cheetah started  at a relative
distance  $y$,  satisfies  the  corresponding  backward  Fokker-Planck
equation:
\begin{equation}
a\,y \frac{dQ(y)}{dy}+2D\frac{d^{2}Q(y)}{dy^{2}}=0\ .
\label{bfp1}
\end{equation}
Since the antelope  is eliminated on meeting the  cheetah, the survival
probability  has  to  vanish  for  $y=0$: $Q(0)=0$.  If  the  prey  is
initially infinitely far from  the predator it will certainly survive,
so   $Q(\infty)=1$.  Solving   the  backward   Fokker-Planck  equation
(\ref{bfp1}) with the stated boundary conditions gives
\begin{equation}
Q(y) = {\rm Erf}\left(\sqrt{\frac{a}{4D}}y\right)\ ,
\end{equation}
where ${\rm Erf}(x)$ is the error function. This result will occur again 
in the next section as a borderline case.


\section{A cheetah surrounded by two antelopes}

In this section we  investigate the infinite-time survival probability
of two antelopes surrounding a  cheetah. To address the problem
in  a   simple  way,  we  interpret   the  individual  one-dimensional
coordinates of  the antelopes  and the cheetah,  $x_{A_{1}}$, $x_{C}$,
$x_{A_{2}}$,  as the  coordinates of  a single  diffusing  particle in
three  dimensions, which  are projected  down  to the  diffusion of  a
single particle in  a two-dimensional absorbing wedge in  the space of
relative   coordinates.  The  boundary   conditions  imposed   by  the
elimination process of the antelopes on meeting the cheetah correspond
to the boundaries of the absorbing wedge.

The  antelopes  and  the  cheetah  evolve according  to  the  Langevin
equation   (\ref{langevin})  with   noise   correlator  (\ref{noise}).
Mapping  this   process  onto  a   single  diffusing  particle   in  a
two-dimensional    wedge,   we    use    the   relative    coordinates
$y_{1}=x_{C}-x_{A_{1}}$  and  $y_{2}=x_{A_{2}}-x_{C}$. This  diffusing
particle now obeys the following equation of motion:
\begin{equation}
\dot{y}_{j}=ay_{j}+\xi _{j},\quad j=1,2,
\end{equation}
where  $\xi_{j}$ is  the `relative'  Gaussian white  noise  defined by
$\xi_{1}=\eta_C-\eta_{A_1}$ and $\xi_{2}=\eta_{A_2}-\eta_C$.  The mean
is zero as beforehand but the correlator now becomes
\begin{displaymath}
 \left<\xi_{i}(t)\xi_{j}(t')\right>=\left\{           \begin{array}{ll}
4D\delta(t-t') & \textrm{for $i=j$}\ ,\\ -2D\delta(t-t') & \textrm{for
$i\neq j$}\ .
\end{array}\right.
\end{displaymath}
Note that exactly the same  equations for the relative coordinates are
obtained   if   the   individual   coordinates  obey   the   equations
$\dot{x}_{A_1} = a(x_C - x_{A_1}) + \eta_{A_1}$, $\dot{x}_C = \eta_C$,
$\dot{x}_{A_2}   =   a(x_{A_2}  -   x_C)   +   \eta_{A_2}$.  In   this
representation, the cheetah is  only diffusing (hence `drowsy'), while
the  antelopes  have  both   diffusive  and  deterministic (`flight') 
components to their motion.

To determine the infinite  time survival probability of the equivalent
single  diffusing   particle  in   two  dimensions  we   consider  the
time-independent  backward  Fokker-Planck   equation  in  the  initial
coordinates $y_{1},y_{2}$:
\begin{eqnarray}
\label{tibfp}
a\left(y_{1}\frac{\partial}{\partial
y_{1}}+y_{2}\frac{\partial}{\partial
y_{2}}\right)Q(y_{1},y_{2})&+&\nonumber\\
2D\left(\frac{\partial^{2}}{\partial y_{1}^{2}}+\frac{\partial^{2}}
{\partial y_{2}^{2}}-\frac{\partial^{2}}{\partial y_{1}\partial
y_{2}}\right)Q(y_{1},y_{2})&=&0.
\end{eqnarray} 

\begin{figure}[htbp]
\centering 
\includegraphics[width=7cm]{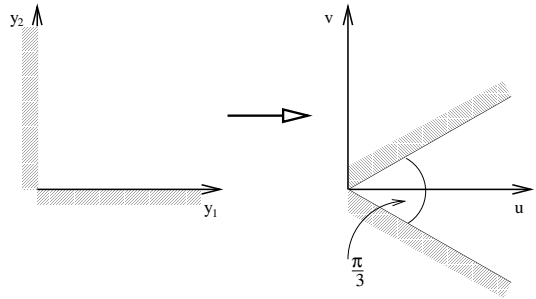}
\caption[Mapping three vicious walkers in an inverted harmonic 
potential to a wedge geometry]
{The transformation to a canonical differential equation maps the 
right-angled wedge in $(y_{1},y_{2})$ coordinates to an axisymmetric 
wedge of opening angle $\Theta=\pi/3$. }
\label{variables}
\end{figure}

Since an antelope  is eliminated on meeting the  cheetah, the survival
probability of the single random  walker must vanish when $y_{1}=0$ or
$y_{2}=0$, corresponding  to the absorbing boundaries of  a wedge with
opening  angle $\Theta=\pi/2$, in  which the  single random  walker is
diffusing,  see   figure  \ref{variables}.   If   both  antelopes  are
infinitely  far from  the cheetah,  the survival  probability  will be
unity, hence $Q(\infty,y_{2}) = Q(y_{1},\infty)=1$.

In  order to  reduce equation  (\ref{tibfp})  to a  canonical form,  a
change  of variables  is required.  The variables  are  first rendered
dimensionless     by    the     change     of    variables     $\tilde
{y}_{i}=y_{i}\sqrt{a/2D}$, $i=1,2$.  Introducing the new variables $u$
and $v$ according to
\begin{equation}
\tilde{y}_{1}=\frac{u+\sqrt{3}v}{2}\quad \tilde{y}_{2}=\frac{u-\sqrt{3}v}{2},
\end{equation}
transforms equation (\ref{tibfp}) to:
\begin{equation}
\left[u \frac{\partial}{\partial u}+v\frac{\partial}{\partial v}+
\frac{\partial^{2}}{\partial u^{2}}+\frac{\partial^{2}}
{\partial v^{2}}\right]Q(u,v)=0.
\end{equation}

The  absorbing boundaries  in the  new variables  $u$ and  $v$  are at
$u=\pm  \sqrt{3}v$. In  the  new variables,  therefore,  the wedge  is
symmetric   about  the   $u$-axis  and   has  an   opening   angle  of
$\Theta=\pi/3$ -- see  figure~\ref{variables}. Because of the symmetry
of the  wedge, polar coordinates $(r,\varphi)$  are appropriate. Hence
the time-independent backward Fokker-Planck equation becomes:
\begin{equation}
\label{pdf3}
\left[\frac{\partial^{2}}{\partial r^{2}}+\frac{1}{r^{2}}
\frac{\partial^{2}}{\partial \varphi^{2}}+\left(\frac{1}{r}+r\right)
\frac{\partial}{\partial r}\right]Q(r,\varphi)=0.
\end{equation}
The  boundary conditions  reduce to  $Q(r,\pi/6) =  Q(r,-\pi/6)=0$ and
$Q(r=0,\varphi)=0$  at  the  absorbing  boundaries of  the  wedge  and
$Q(\infty,\varphi)=1$ for $-\pi/6 <  \varphi < \pi/6$ corresponding to
the  survival of  both antelopes  if  they are  initially at  infinite
distance from the cheetah.

The partial differential equation (\ref{pdf3}) can be solved by 
separation of variables,
\begin{equation}
Q(r,\varphi)=\sum_{n=1}^{\infty}A_{n}R_{n}(r)\Phi_{n} (\varphi)\ ,
\label{Qexpansion}
\end{equation}
where the angular part $\Phi_{n}( \varphi)$ is a cosine mode satisfying 
the angular boundary conditions, 
\begin{equation}
\label{angular}
\Phi_{n}(\varphi)=\cos(3(2n-1)\varphi)\ ,
\end{equation}
and the coefficients $A_{n}$ are to be determined by the radial boundary 
conditions.

Substituting the result for $\Phi_{n}(\varphi)$ in  (\ref{pdf3}) yields 
the following ordinary differential equation for $R_{n}(r)$.
\begin{equation}
r^{2}R_{n}''(r)+\left(r+r^{3}\right)R_{n}'(r)-9(2n-1)^{2}R_{n}(r)=0\ .
\end{equation}
By setting $r^{2}=\zeta$ and $R_{n}(r)=\zeta^{3n-\frac{3}{2}} \rho_{n} 
(\zeta )$ this differential equation is transformed into 
\begin{equation}
\label{rho}
\zeta \rho_{n}''(\zeta)+\left(\frac{1}{2}\zeta+6n-2\right)\rho_{n}'
(\zeta)+\frac{6n-3}{4}\rho_{n}(\zeta)=0.
\end{equation}
This ordinary differential equation is related to the confluent 
hypergeometric differential equation (see 2.273(9) in 
reference~\cite{kamke}). Defining $\zeta=2\sigma $
and $\rho_{n}(\zeta)=\exp(-\sigma)\psi_{n}(\sigma)$, equation (\ref{rho}) 
reduces to the confluent hypergeometric differential equation, also
called Kummer's equation~\cite{kamke, abramowitz}
\begin{equation}
\sigma\psi_{n}''(\sigma)+(6n-2-\sigma)\psi_{n}'(\sigma)
-\left(3n-\frac{1}{2}\right)\psi_{n}(\sigma)=0.
\end{equation}
The solutions of this differential equation are known. The general 
solution can be written in terms of Kummer's function of the first 
kind, ${\rm M}(a,b,z)$, and of the second kind, ${\rm U}(a,b,z)$, also 
denoted confluent hypergeometric functions of the first and second kind 
\cite{abramowitz}:
\begin{eqnarray}
\psi_{n}(\sigma)&=&B_{n}{\rm M}\left(3n-\frac{1}{2},6n-2,\sigma\right)
\nonumber\\
&+&C_{n}{\rm U}\left(3n-\frac{1}{2},6n-2,\sigma\right),
\end{eqnarray}
where  $B_{n}$ and  $C_{n}$  are  constants to  be  determined by  the
boundary  condition.   Note  that   we  have  introduced,   for  later
convenience, a redundancy in the coefficients, having $A_n$, $B_n$ and
$C_n$ when there  are only two independent sets  of coefficients. This
redundancy  will  be  removed  below  by an  explicit  choice  of  the
coefficients $B_n$.

Substituting all former transformations, the result for $R_{n}(r)$ is
\begin{eqnarray}
R_{n}(r)& = & B_n\,r^{6n-3}e^{-\frac{r^{2}}{2}}\, 
{\rm M}\left(3n-\frac{1}{2},6n-2,\frac{r^{2}}{2}\right)\nonumber\\ 
& + & C_n\,r^{6n-3}e^{-\frac{r^{2}}{2}}\,
{\rm U}\left(3n-\frac{1}{2},6n-2,\frac{r^{2}}{2}\right).
\end{eqnarray}
The particular solution we are looking  for has to vanish at $r=0$ and
approach  a constant value  for $r\rightarrow  \infty$ to  satisfy the
boundary  conditions.  The confluent  hypergeometric  function of  the
first kind  is unity  when its argument  is zero,  ${\rm M}(a,b,0)=1$,
whereas  the  hypergeometric  function   of  the  second  kind,  ${\rm
U}(a,b,z)$, diverges as $z \to 0$ for $b>1$ \cite{abramowitz}
which is the case in our solution, where $b=6n-2$, since $n>0$. 
Hence we set $C_{n}=0$ in the solution so that it vanishes at $r=0$.

Now we investigate the behaviour of our solution in the limit 
$r\rightarrow \infty$. The asymptotic form of the hypergeometric
function of the first kind for large arguments, $z\rightarrow +\infty$, 
is \cite{abramowitz}:
\begin{equation}
{\rm M}(a,b,z) \sim \frac{\Gamma (b)}{\Gamma (a)}z^{a-b}e^z\ .
\end{equation}
Hence the radial solution approaches a constant value for 
$r\rightarrow\infty$.
\begin{equation}
\label{limit}
\lim_{r\rightarrow \infty}R_{n}(r)=2^{3n-\frac{3}{2}}
\frac{\Gamma(6n-2)}{\Gamma(3n-1/2)}\,B_n
\end{equation}
To    simplify    the    fitting    to    the    boundary    condition
$Q(r=\infty,\varphi)=1$ we eliminate  the aforementioned redundancy in
the expansion coefficients by choosing the constants $B_{n}$ such that
$R_n(\infty)=1$   for    all   $n$,    i.e.\   we   choose    $B_n   =
2^{3/2-3n}\frac{\Gamma(3n-1/2)}{\Gamma(6n-2)}$.     The   coefficients
$A_{n}$ in  Eq.\ (\ref{Qexpansion}) can be determined  by imposing the
boundary  condition $Q(\infty,\phi)=1$,  i.e.\  $\sum_{n=1}^\infty A_n
\cos[3(2n-1)\phi]=1$,      for     $\phi$     in      the     interval
$(-\pi/6,\pi/6)$. This gives
\begin{equation}
A_{n}=\frac{4}{\pi}\frac{(-1)^{n-1}}{2n-1}\ .
\end{equation}

Finally we  simplify the  radial solution by  use of  Kummer's formula
\cite{abramowitz}:
\begin{equation}
e^{z}{\rm M}(a,b,-z)={\rm M}(b-a,b,z)\ .
\end{equation}
Then the  solution for the  infinite time survival probability  of the
single  diffusing particle in  a wedge  becomes, in  the dimensionless
variables $(r,\varphi)$,
\begin{eqnarray}
Q(r,\varphi)&=&\sum_{n=1}^{\infty}2^{-3n+\frac{7}{2}}
\frac{\Gamma(3n-1/2)}{\pi(2n-1)\Gamma(6n-2)}\nonumber\\
&\times& (-1)^{n-1} \cos(3(2n-1)\varphi)\nonumber\\
&\times& r^{6n-3}{\rm M}\left(3n-\frac{3}{2},6n-2,-\frac{r^{2}}{2}\right).
\end{eqnarray}
This sum is easily shown  to converge since the summand $a_{n}$ decays
to zero faster than $1/n$ for $n\rightarrow\infty$. For large $n$, the
confluent  hypergeometric  function  approaches exponential  function,
${\rm      M}\left(3n-\frac{3}{2},6n-2,-\frac{r^{2}}{2}\right)     \to
\rm{exp}(-r^{2}/4)$.   The asymptotic  form of  the quotient  of gamma
functions    is    given    by    $\Gamma(3n-1/2)/\Gamma(6n-2)    \sim
2^{-3n+1}(6n-3)^{-3n+3/2}e^{3n-3/2}$.  In  summary, the summand decays
to zero for large $n$ as
\begin{equation}
a_{n} \sim \frac{2^{-6n+9/2}}{(2n-1)\pi}(6n-3)^{-3n+3/2}r^{6n-3}
e^{3n-3/2-r^{2}/4},
\end{equation}
where the alternating signs and oscillating cosine functions have been
omitted.  Although  the   sum  clearly  converges,  the  computational
equipment was not sufficient to calculate the sum in general.
Therefore, all plots of the solution to be displayed in this paper are
approximations  including the  first 30  terms  of the  sum, which  is
sufficient in the  chosen range, since, for example,  the error due to
the absence  of the  next ten terms,  up to  term 40, is  smaller than
$5\times10^{-37}$.

\begin{figure}[htb]
\hspace*{0.3cm}{\input{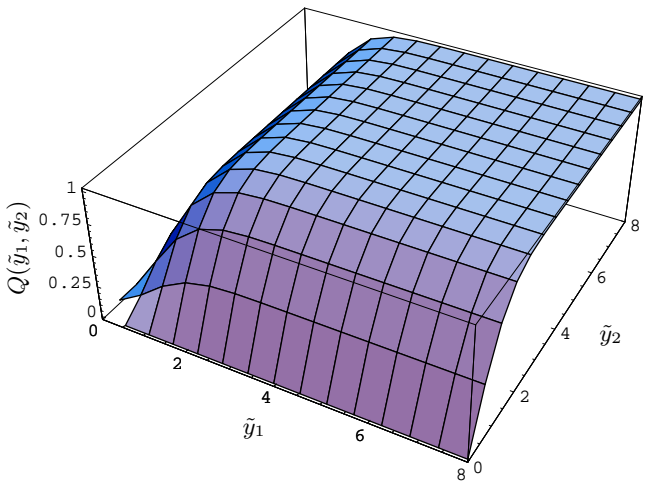}}
\caption[The infinite time survival probability of two antelopes 
surrounding a  cheetah]
{The infinite-time survival probability of two antelopes surrounding a 
cheetah, plotted against the dimensionless relative coordinates
$\tilde{y}_{1}=\sqrt{a/2D}(x_{2}-x_{1})$ and 
$\tilde{y}_{2}=\sqrt{a/2D}(x_{3}-x_{2})$. }
\label{2010}
\end{figure}

To  plot  and  analyse  the  infinite  time  survival  probability  we
transform the solution back  to the dimensionless relative coordinates
$\tilde{y}_{1}$ and  $\tilde{y}_{2}$. In those  coordinates the result
reads:
\begin{eqnarray}
\label{solyy}
Q(\tilde{y}_{1},\tilde{y}_{2})&=&\sum_{n=1}^{\infty}(-1)^{n-1}
2^{3n+\frac{1}{2}}\frac{\Gamma(3n-1/2)}{\pi(2n-1)\Gamma(6n-2)}\nonumber\\
&\times&\cos\left[3(2n-1)\arctan\left(\frac{\tilde{y}_{1}-\tilde{y}_{2}}
{\sqrt{3}(\tilde{y}_{1}+\tilde{y}_{2})}\right)\right]\nonumber\\
&\times&{\rm M}\left(3n-\frac{3}{2},6n-2,-\frac{2}{3}(\tilde{y}_{1}^{2}
+\tilde{y}_{1}\tilde{y}_{2}+\tilde{y}_{2}^{2})\right)\nonumber\\
&\times& \left(\frac{1}{3}(\tilde{y}_{1}^{2}+\tilde{y}_{1}\tilde{y}_{2}
+\tilde{y}_{2}^{2})\right)^{3n-3/2}.
\end{eqnarray}
In figure~\ref{2010} this function is plotted in the range 
$\tilde{y}_{1},\tilde{y}_{2}\in[0,8]$.
The  survival probability smoothly  increases from  zero on  the lines
$\tilde{y}_{1}=0$ and  $\tilde{y}_{2}=0$ to  form a plateau  of almost
constant probability for  $\tilde{y}_{1}>2$ and $\tilde{y}_{2}>2$ that
increases      to     unity     at      $\tilde{y}_{1}=\infty$     and
$\tilde{y}_{2}=\infty$,  corresponding to  certain survival  when both
antelopes  start  infinitely  far  from  the  cheetah.   Unfortunately
Mathematica could not  calculate the sum for $\tilde{y}_{1}\rightarrow
0$  and  $\tilde{y}_{2}\rightarrow 0$,  but  the  summand of  equation
(\ref{solyy})   clearly   vanishes    when   $\tilde{y}_{1}=   0$   or
$\tilde{y}_{2}= 0$ due to the vanishing of the cosine functions.

\begin{figure}[ht]
\hspace*{1cm}{\input{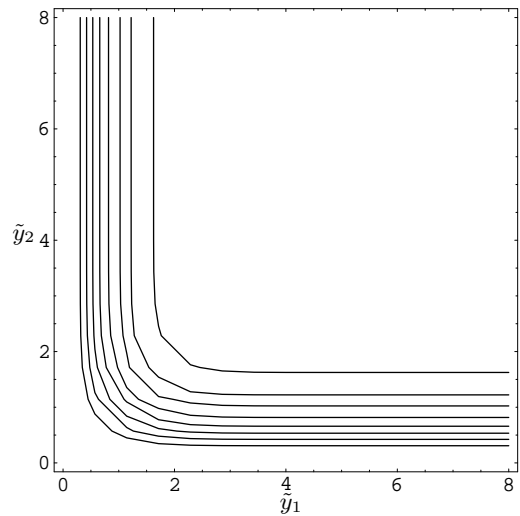}}
\caption[Contour lines of the infinite time survival probability of 
two antelopes surrounding a cheetah]
{Contour lines of the infinite time survival probability of two 
antelopes surrounding a  cheetah versus the relative coordinates 
$\tilde{y}_{1}=\sqrt{a/2D}(x_{2}-x_{1}),$ and 
$\tilde{y}_{2}=\sqrt{a/2D}(x_{3}-x_{2})$. 
The different lines correspond to constant probabilities of 0.1 up to 0.8.}
\label{contour2010}
\end{figure}

To study the  survival probability further, it is  also of interest to
consider  the   contour  lines   of  figure~\ref{2010}  as   shown  in
figure~\ref{contour2010}.
Investigating  those  one  easily  recognises  that  the  function  is
symmetric about the line $\tilde{y}_{1}=\tilde{y}_{2}$, as it must be.
Furthermore,  in  the limit  of  one  relative  coordinate tending  to
infinity, say  $\tilde{y}_{2}=\infty$, the problem  with two antelopes
simplifies to the  problem of a single antelope  with a cheetah, which
has   been   calculated   in   section~\ref{simpleexample}.   In   the
dimensionless variables, the result  for the survival probability of a
single antelope and a cheetah is:
\begin{equation}
Q(\tilde{y}_{1},\infty)={\rm Erf}\left(\frac{\tilde{y}_{1}}{\sqrt{2}}\right).
\label{error2}
\end{equation}
\begin{figure}[htb]
\hspace*{1.3cm}{\input{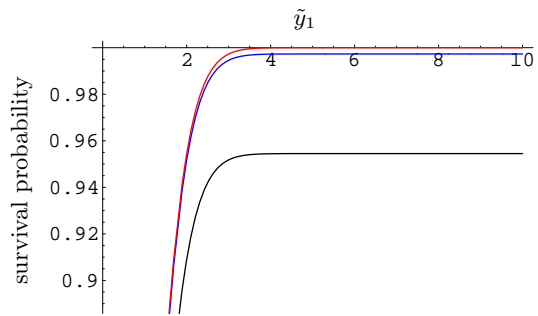}}
\caption[The infinite-time survival probability of two antelopes 
surrounding a cheetah keeping one relative coordinate fixed ]
{The infinite time survival probability of two antelopes surrounding 
a cheetah keeping the relative coordinate $\tilde{y}_{2}=c$ fixed at 
(bottom to top) $c=2$, $c=3$ and $c=4$, where the top curve is already 
indistinguishable from the error function (\ref{error2}).} 
\label{error}
\end{figure}
Unfortunately,  extracting this  limiting  behaviour analytically  has
proved      to      be      intractable.     Instead      we      plot
$Q(\tilde{y}_{1},\tilde{y}_{2}=c)$       for       $c=2,3,4$,      see
figure~\ref{error}.  The  figure clearly  shows  how  the sequence  of
curves     approaches    the     error    function     expected    for
$\tilde{y}_{2}=\infty$, see equation  (\ref{error2}).  The $c=4$ curve
lies  on  top  of  the  error  function,  demonstrating  the  limiting
behaviour.

\section{Conclusion}
In this  paper we  introduced the interesting  problem of  a diffusion
controlled reaction  where, in addition  to the diffusive  motion, the
particles are subjected to a separating drift.  By mapping the process
of two antelopes  surrounding a cheetah to that  of a single diffusing
particle  in two  dimensions,  we derived  the  probability that  both
antelopes have  survived up  to infinite time  as a function  of their
initial separations from the cheetah.


\begin{thebibliography}{30}
\expandafter\ifx\csname natexlab\endcsname\relax\def\natexlab#1{#1}\fi
\expandafter\ifx\csname bibnamefont\endcsname\relax
  \def\bibnamefont#1{#1}\fi
\expandafter\ifx\csname bibfnamefont\endcsname\relax
  \def\bibfnamefont#1{#1}\fi
\expandafter\ifx\csname citenamefont\endcsname\relax
  \def\citenamefont#1{#1}\fi
\expandafter\ifx\csname url\endcsname\relax
  \def\url#1{\texttt{#1}}\fi
\expandafter\ifx\csname urlprefix\endcsname\relax\def\urlprefix{URL }\fi
\providecommand{\bibinfo}[2]{#2}
\providecommand{\eprint}[2][]{\url{#2}}

\bibitem[{\citenamefont{fisher and gelfand}(1988)}]{fishergelfand}
\bibinfo{author}{\bibfnamefont{M.~E.} \bibnamefont{Fisher}} \bibnamefont{and}
  \bibinfo{author}{\bibfnamefont{M. ~P.} \bibnamefont{Gelfand}},
  \bibinfo{journal}{J. Stat. Phys.}\textbf{\bibinfo{volume}{53}},
  \bibinfo{pages}{175}
  (\bibinfo{year}{1988}).

\bibitem[{\citenamefont{ben-avraham}(1984)}]{benavraham}
\bibinfo{author}{\bibfnamefont{D.} ~\bibnamefont{ben Avraham}},
  \bibinfo{journal}{J. Chem. Phys.}\textbf{\bibinfo{volume}{88}},
  \bibinfo{pages}{941}
  (\bibinfo{year}{1988}).

\bibitem[{\citenamefont{krapivsky and redner}(1996)}]{krapivskyredner96}
\bibinfo{author}{\bibfnamefont{K. ~L.} \bibnamefont{Krapivsky}} 
\bibnamefont{and}
\bibinfo{author}{\bibfnamefont{S.} ~\bibnamefont{Redner}},
  \bibinfo{journal}{J. Phys. A} \textbf{\bibinfo{volume}{29}},
  \bibinfo{pages}{5347}
  (\bibinfo{year}{1996}).

\bibitem[{\citenamefont{redner and krapivsky}(1999)}]{rednerkrapivsky99}
\bibinfo{author}{\bibfnamefont{S.} ~\bibnamefont{Redner}} \bibnamefont{and}
\bibinfo{author}{\bibfnamefont{K. ~L.} \bibnamefont{Krapivsky}},
  \bibinfo{journal}{Am. J. Phys.} \textbf{\bibinfo{volume}{67}},
  \bibinfo{pages}{1277}
  (\bibinfo{year}{1999}).


\bibitem[{\citenamefont{BW}(2004)}]{BW}
\bibinfo{author}{\bibfnamefont{A. ~J.} ~\bibnamefont{Bray}} 
\bibnamefont{and}
\bibinfo{author}{\bibfnamefont{K.} ~\bibnamefont{Winkler}},
  \bibinfo{journal}{J. Phys. A}\textbf{\bibinfo{volume}{37}},
  \bibinfo{pages}{5493}
  (\bibinfo{year}{2004}).


\bibitem[{\citenamefont{kamke}(1977)}]{kamke}
\bibinfo{author}{\bibfnamefont{E.} ~\bibnamefont{Kamke}},
  \emph{\bibinfo{title}{Differentialgleichungen: L\"osungsmethoden und L\"osungen}}
  (\bibinfo{publisher}{B. G. Teubner}, \bibinfo{address}{Stuttgart},
  \bibinfo{year}{1977}, vol. 1, chapter C2.

\bibitem[{\citenamefont{abramowitz and stegun}(1972)}]{abramowitz} 
\bibinfo{author}{\bibfnamefont{M.} ~\bibnamefont{Abramowitz}}
\bibnamefont{and}
\bibinfo{author}{\bibfnamefont{I. ~A.} \bibnamefont{Stegun}},
  \emph{\bibinfo{title}{Handbook of Mathematical Functions}}
  (\bibinfo{publisher}{Dover}, \bibinfo{address}{New York},
  \bibinfo{year}{1972}), chapter 13.



\end{thebibliography}
\end{document}